\documentclass{aa}  

\usepackage{csquotes}
\usepackage{graphicx}
\usepackage{caption}
\usepackage{txfonts}
\usepackage{hyperref}
\begin{document}

   \title{Star formation around three co-moving HAeBe stars in the Cepheus Flare}

    \subtitle{}
    \titlerunning{Star formation around three co-moving HAeBe stars}
    \authorrunning{Piyali Saha et. al}

   \author{Piyali Saha
          \inst{1,2}
          \and
          Maheswar G.\inst{1}
          \and
          Blesson Mathew\inst{3}
          \and
          U. S. Kamath\inst{1}
          }

   \institute{Indian Institute of Astrophysics (IIA), Sarjapur Road, Koramangala, Bangalore 560034, India\\
        \email{s.piyali16@gmail.com}
            \and
             Pt. Ravishankar Shukla University, Amanaka G.E. Road, Raipur, Chhatisgarh 492010, India
             \and
             Department of Physics and Electronics, CHRIST (Deemed to be University), Bangalore 560029, India}

   \date{Received ....; accepted ....}

 
  \abstract
   {The presence of three more Herbig Ae/Be (HAeBe) candidates in the Cepheus Flare within a 1.5$\degr$ radius centered on HD 200775 suggests that star formation is prevalent in a wider region of the LDN 1147/1158, LDN 1172/1174, and LDN 1177 clouds. A number of young stellar objects (YSOs) are found to be distributed toward these cloud complexes along with the HAeBe stars. Various star formation studies clearly indicate ongoing low-mass star formation inside the clouds of this region. Sources associated with less near-infrared excess and less H$\alpha$ emission raise the possibility that more low-mass YSOs, which were not identified in previous studies, are present in this region.}
   {The aim is to conduct a search for additional young sources that are kinematically associated with the previously known YSOs and to characterize their properties.}
   {Based on the \textit{Gaia} DR2 distances and proper motions, we found that the HAeBe candidates BD+68$\degr$1118, HD 200775, and PV Cep are all spatially and kinematically associated with previously known YSOs. Based on the \textit{Gaia} DR2 data, we identified a number of co-moving sources around BD+68$\degr$1118. These sources are characterized using optical and near-infrared color-color and color-magnitude diagrams.}
   {We estimated a distance of $340\pm7$ pc to the whole association that contains BD+68$\degr$1118, HD 200775, and PV Cep. Based on the distance and proper motions of all the known YSOs, a total of 74 additional co-moving sources are found in this region, of which 39 form a loose association surrounding BD+68$\degr$1118. These sources are predominantly M-type sources with ages of $\sim$10 Myr and no or very little near-infrared excess emission. The distribution of co-moving sources around BD+68$\degr$1118 is much more scattered than that of sources found around HD 200775. The positive expansion coefficients obtained via the projected internal motions of the sources surrounding BD+68$\degr$1118 and HD 200775 show that the co-moving sources are in a state of expansion with respect to their HAeBe stars. A spatio-temporal gradient of these sources toward the center of the Cepheus Flare Shell supports the concept of star formation triggered by external impacts.}
   {}


   \maketitle
%

\section{Introduction} \label{sec:intro}

    It is now widely accepted that a majority of star formation in the Galaxy does not occur in isolation but happens in groups or clusters \citep[e.g., ][]{1985ApJ...294..523E, 1999ARA&A..37..311E,  2000AJ....120.3139C, 2003ARA&A..41...57L}. These aggregations are characterized by the number of members (N) contained in them and are often classified as groups if N $<$ 100 and clusters if N $>$ 100 \citep[e.g., ][]{2001ApJ...553..744A}. The richness of a stellar system depends on its most massive member, with low-mass stars forming in small loose aggregates and high-mass stars usually found in dense clusters \citep{2007ARA&A..45..481Z}. There is a growing body of evidence that even the Sun was formed in a group or cluster of stars, with a high-mass star that exploded as a supernova while our Sun was in its early stage of formation \citep[e.g., ][]{2006ApJ...652.1755L, 2010ARA&A..48...47A}. The intermediate-mass stars falling in the mass interval $2\lesssim {M/M}_{\sun} \lesssim 10$ are therefore of particular interest as they form a connection between well-understood low-mass star formation and the enigmatic high-mass star formation. The pre-main-sequence (PMS) stars that fall in the above mass range are known as Herbig Ae/Be (HAeBe) stars \citep{1960ApJS....4..337H, 1998ARA&A..36..233W}. First categorized by \cite{1960ApJS....4..337H}, HAeBe stars are of spectral types B, A, and F, show emission lines, are located in an obscured region, and illuminate bright nebulas in their immediate vicinities. However, later observations revealed that these types of sources can be found in isolation and show infrared (IR) excess due to the presence of circumstellar disks \citep{1989A&A...208..213H, 1992A&AS...96..625O, 1994A&AS..104..315T, 1998ARA&A..36..233W, 2000A&A...357..325V, 2001A&A...365..476M}. Several astrometric and spectroscopic studies have recently been carried out on a large number of HAeBe stars compiled from the literature \citep{2015MNRAS.453..976F,2018A&A...620A.128V,2019AJ....157..159A, 2020MNRAS.493..234W}. 
    
    Several photometric studies of the fields containing HAeBe stars were performed in near-IR \citep{1997A&A...320..159T, 1998A&AS..133...81T, 1999A&A...342..515T, 2007ApJ...659.1360W} and mid-IR \citep{2003A&A...400..575H} bands. These studies were based on the notion that the reduced extinction at these wavelength regimes would allow the detection of accompanying populations of low-mass embedded young stellar objects (YSOs) that might have possibly formed along with HAeBe stars in the same star formation event. The results suggested an apparent relationship between the spectral type of HAeBe stars and the richness of the embedded population around them. While early-type Be stars are usually found within rich clusters, late-type Be and Ae stars are never associated with any discernible group of YSOs \citep{1997A&A...320..159T, 1998A&AS..133...81T, 1999A&A...342..515T}.   

    The low-mass counterparts of HAeBe stars are T Tauri stars (TTSs): YSOs that show Balmer lines of hydrogen in emission \citep{1989A&ARv...1..291A}, which is considered a sign of active accretion of material from the surrounding circumstellar disk. Based on the equivalent width (EW) of H${\alpha}$ emission lines, TTSs are classified into classical T Tauri stars (CTTSs) and weak-line T Tauri stars (WTTSs). Classical T Tauri stars show a relatively high EW of H${\alpha}$ emission due to the accretion of circumstellar material. In WTTSs, the H${\alpha}$ is either weakly present -- with EWs typically $\lesssim$10 \AA,~which is considered to originate from the chromosphere \citep{2003AJ....126.2997B} -- or absent (naked TTSs). Based on an empirical sequence, YSOs can also be classified using the slope of their spectral energy distribution (SED) in the near-IR to submillimeter wavelengths. The low-mass YSOs that are deeply embedded are classified as \textquote{Class 0} objects, whereas those that are more evolved but still embedded in their envelope are \textquote{Class I} objects. Sources showing excess IR emission due to the presence of circumstellar material in a flattened geometry are \textquote{Class II} objects, and sources having little or no IR excess emission are \textquote{Class III} objects (\citealp{1987IAUS..115....1L}; \citealp{1993ApJ...406..122A}; \citealp{1994ApJ...434..614G}). Though refinements to the above classification scheme were suggested in subsequent studies \citep[e.g., ][]{2007ApJS..169..328R, 2009ApJS..181..321E, 2010ApJS..188...75M}, it adequately presents an evolutionary sequence of YSOs. The Class II and Class III objects generally correspond to CTTSs and WTTSs, respectively. Thus, low-mass YSOs can be identified by carrying out photometry in near- and mid-IR bands as well as with H${\alpha}$ surveys to detect emission line sources. Acquiring the physical properties of high-mass HAeBe stars and low-mass TTSs is necessary for a full understanding of their formation and evolution.

    Of the HAeBe stars studied by \citet{1998A&AS..133...81T}, HD 200775 was found to be peculiar as no surface density enhancement was detected despite it being a B2/3Ve spectral type star \citep{1994A&AS..104..315T,2006ApJ...653..657M}. HD 200775 is responsible for illuminating a bright reflection nebula, NGC 7023. This nebula is located at the northern edge of an elongated molecular cloud, LDN 1172/1174 \citep[hereafter L1172/1174; ][]{1962ApJS....7....1L}, situated at a relatively high Galactic latitude ($l\sim104.1\degr$, $b\sim+14.2\degr$) in the Cepheus Flare region at a distance of 335$\pm$11 pc \citep[][hereafter Paper I]{2020MNRAS.494.5851S}. \cite{2009ApJS..185..451K} conducted a comprehensive study of star formation toward the Cepheus Flare region, in which several signposts of low- to intermediate-mass star formation were analyzed. Another cloud complex, LDN 1147/1158 (hereafter L1147/1158), located about $2^{\circ}$ to the west of L1172/1174, is also found to be located at a distance of 340$\pm$3 pc \citep{2020A&A...639A.133S}, consistent with that of L1172/1174. In addition to these, LDN 1177 (hereafter L1177), also known as CB 230 \citep{1988ApJS...68..257C}, is located to the east of L1172/1174 at an angular distance of $1.5^{\circ}$. One of its stars, BD+67$\degr$1300, which illuminates the reflection nebula GN 21.15.8 \citep{2003A&A...399..141M} and is hence associated with L1177 \citep{2009ApJS..185..451K}, is found to be at a distance of 341$^{+2}_{-3}$ pc \citep{2018AJ....156...58B}. The large-scale $^{13}$CO (J$=$1$-$0) survey of the Cepheus and Cassiopia region made by \citet{1997ApJS..110...21Y} showed that these clouds share similar radial velocities ($2.7-2.9$ km~s$^{-1}$), suggesting that all three regions are both spatially and kinematically connected.  
 
\begin{figure}
        \centering
    \includegraphics[height=8cm, width=9.3cm]{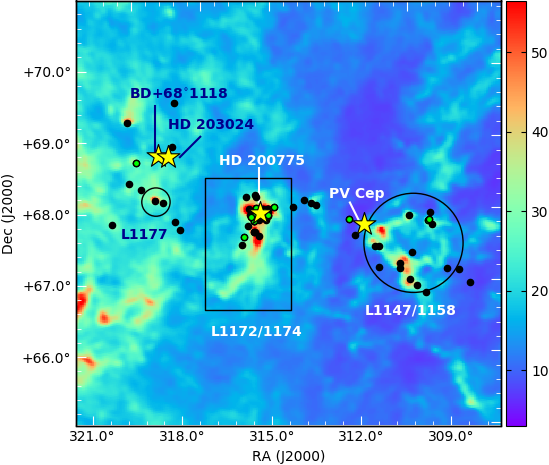}
    \caption{Area studied in this work, shown in a 5$\degr\times$5$\degr$ \textit{Planck} 857 GHz image. The regions include the L1147/1158, L1172/1174, and L1177 cloud groups. The circles and the rectangle show the extent of these groups. Positions of four intermediate-mass stars, HD 203024, BD+68$\degr$1118, HD 200775, and PV Cep, are indicated by yellow star symbols. The black filled circles show the locations of YSO candidates obtained from the literature and not detected in \textit{Gaia} DR2. The green filled circles indicate the YSO candidates that have reliable detection in \textit{Gaia} DR2.}
    \label{fig:cepheus_pol_YSO}
\end{figure}

    PV Cep, associated with the reflection nebula GM 29 \citep{1977SvAL....3...58G}, is an irregular eruptive protostar \citep{1977ApJ...215L.127C} situated at the northeastern edge of the L1147/1158 complex. As PV Cep is highly variable, its nature is not very clear, although it has been studied extensively at different wavelengths.
    \cite{1981ApJ...245..920C} estimated the spectral type of PV Cep to be A5 based on the Balmer absorption lines, which were absent after the 1977$-$1979 outburst. It could be possible that the spectrum represented only a shell spectrum \citep{1981ApJ...245..920C}. Later, an absorption spectrum similar to that of a G8-K0 spectral type was reported by \citet{2001Ap.....44..419M}. Additionally, the combination of the presence of H$_{2}$O MASER emission, a relatively massive circumstellar disk, a spectrum showing UV pumped lines in emission, and a high jet velocity supports the conjecture that PV Cep is an embedded and young Herbig Ae star \citep{2013A&A...554A..66C}. Two emission line stars, HD 203024 and BD+68$^{\circ}$1118, are found to be located in the north, toward the outer edge of the diffuse part of L1177. Both the stars, which are of B8.5V spectral type, were classified as HAeBe candidates by \citet{2000MNRAS.319..777K} and are considered to be part of the region. Later, \cite{2009ApJS..185..451K} estimated the spectral type of BD+68$^{\circ}$1118 to be A2. The distance and the proper motion values of BD+68$\degr$1118 ($d=334^{+2}_{-3}$ pc, ~$\mu_{\alpha\star}=7.778\pm0.042$ mas yr$^{-1}$, ~$\mu_{\delta}=-0.992\pm0.040$ mas yr$^{-1}$) and PV Cep ($d=341^{+7}_{-7}$ pc, $\mu_{\alpha}=8.228\pm0.126$ mas yr$^{-1}$, $\mu_{\delta}=-1.976\pm0.110$ mas yr$^{-1}$) obtained from the second \textit{Gaia} data release (\textit{Gaia} DR2) are found to be consistent with those of L1172/1174, as shown in Paper I. Thus, there are at least three early-type sources, namely HD 200775, BD+68$^{\circ}$1118, and PV Cep (there is no \textit{Gaia} DR2 data for HD 203024), associated with the three star forming regions. These results imply that star formation is prevalent around L1172/1174 and extends over a wider area. The molecular cloud groups L1147/1158, L1172/1174, and L1177, the intermediate-mass stars PV Cep, HD 200775, BD+68$\degr$1118, and HD 203024, and the known YSO candidates found toward the region are shown in Fig. \ref{fig:cepheus_pol_YSO}. Paper I presents the identification of 20 new co-moving sources around HD 200775. The question remains as to whether there are new co-moving sources associated with BD+68$\degr$1118 and PV Cep as well. A majority of new co-moving sources found around HD 200775 are low-mass stars that have considerably low near- and mid-IR excess emission. Spectroscopy of some sources shows H$\alpha$ in emission but with smaller EWs ($\sim$10 \AA). Consequently, they escape detection in large-scale H$\alpha$, near-IR, and mid-IR surveys. Therefore, it is possible that co-moving sources could exist around the neighboring HAeBe stars as well. In this study, we searched for this category of sources and conducted a preliminary study of them.

    The investigation for additional co-moving sources was made toward a circular region of radius 3.5$\degr$ centered on HD 200775 that contains the cloud groups L1147/1158, L1172/1174, and L1177, using \textit{Gaia} DR2. Based on optical and IR color-color (CC) diagrams and color-magnitude diagrams (CMDs), we describe their characteristics. A complete census of young sources associated with a region is important since knowledge of their spatial distribution is key to understanding its star formation history. This paper is organized as follows. We present the archival data used in this work in Sect.\ \ref{sec:obs_data}, discuss our results in Sect. \ref{sec:dis}, and conclude with a summary in Sect. \ref{sec:sum_con}.
    
\section{Archival data}\label{sec:obs_data}


\textit{Gaia} DR2 \citep{2018A&A...616A...1G} presents the positions, parallaxes, and proper motions of more than a billion objects with unprecedented precision. However, if the relative uncertainties in parallax values are $\gtrsim$ 20\%, the corresponding distances will not follow the simple inversion of their parallaxes \citep{2015PASP..127..994B}. Recently, \cite{2018AJ....156...58B} provided a probabilistic estimate of the stellar distances from the parallax measurements (provided by \textit{Gaia} DR2) using an exponentially decreasing space density prior, which is based on a galactic model. In our analysis, we obtained the stellar distances and proper motion values from \cite{2018AJ....156...58B} and from \cite{2018A&A...616A...1G}, respectively, by searching within a radius of 1$\arcsec$ around the source positions. 

To characterize the properties of the co-moving sources identified in this work, we obtained their 2MASS \citep{2006AJ....131.1163S} and \textit{WISE} \citep{2010AJ....140.1868W} magnitudes from the \cite{2003yCat.2246....0C} and \cite{2012yCat.2311....0C} catalogs, respectively. We also acquired Pan-STARRS photometric data from \cite{2016arXiv161205560C}. Only sources with photometric quality "A" (S/N$\geq$10) in all bands were considered.

\section{Results and discussion}\label{sec:dis}

\subsection{Search for co-moving sources around BD+68$\degr$1118, HD 200775, and PV Cep}

\begin{figure}
    \centering
    \includegraphics[height=6cm, width=8.4cm]{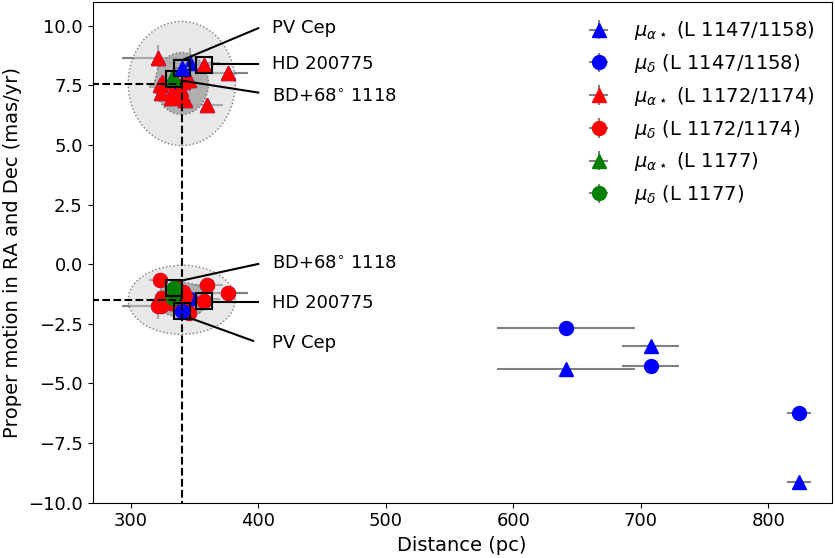}
    \caption{Proper motion values of the known YSO candidates as a function of their distances, obtained from \textit{Gaia} DR2. The triangles and circles in blue, red, and green represent the $d-\mu_{\alpha\star}$ and $d-\mu_{\delta}$ values of the sources located toward L1147/1158, L1172/1174, and L1177, respectively. The error ellipses corresponding to three times the MAD (darker shade) and six times the MAD (lighter shade) in proper motion and distance values are drawn. The locations of HD 200775, BD+68$\degr$1118, and PV Cep are also marked. The dashed lines show the median values of $d$, $\mu_{\alpha\star}$, and $\mu_{\delta}$.}
    \label{fig:pm_dist_yso_cepheus}
\end{figure}

In Paper I, a total of 58 YSO candidates were identified in the vicinity of L1172/1174. In addition, we also found 22 and 13 sources in the vicinities of L1147/1158 and L1177, respectively \citep[][and references in Paper I]{1998ApJS..115...59K,2008ApJS..179..249D}. Applying the selection criteria described in Paper I, we obtained \textit{Gaia} DR2 counterparts for 24 sources (seven from L1147/1158, fifteen from L1172/1174, and two from L1177), well within a search radius of 1$\arcsec$, that had renormalized unit weight errors \citep[RUWEs\footnote{RUWE values are obtained from \url{http://gaia.ari.uni-heidelberg.de/}};][]{LL:LL-124} $\leqslant$1.5. Of the 24 sources, 18 have RUWE$\leqslant$1.4 and 4 have 1.4$<$RUWE$\leqslant$1.5. To also include the latter sources in our analysis, we adopted the limiting value of RUWE to be 1.5 instead of the typical value of 1.4 \citep{LL:LL-124} as one of our criteria for selecting the sources. Though the RUWE value for HD 200775 is $\sim$1.6, we also included it in our analysis. One source, \textit{Gaia} DR2 2270941147188904704, associated with L1147/1158, has no estimated RUWE, nor does it have $G_{\mathrm{BP}}$ or $G_{\mathrm{RP}}$ values in the \textit{Gaia} DR2 catalog; however, it has reliable proper motion and distance measurements (m/$\sigma_{m}\geqslant$ 3) and hence is included in our analysis. Table \ref{tab:YSO_gaia_cepheus} in the appendix lists the YSO candidates toward L1147/1158 and L1177. The 15 YSO candidates toward L1172/1174 are already listed in Table 3 of Paper I, with the exception of sources \#6, \#12, \#16, \#17, and \#18; these 15 sources are therefore not shown in the present paper. We obtained the medians and median absolute deviations (MADs) of $d$, $\mu_{\alpha\star}$, and $\mu_{\delta}$ for these 24 YSO candidates. The medians are $d$ = 340$\pm$7 pc, $\mu_{\alpha\star}$ = 7.580$\pm$0.434 mas yr$^{-1}$, and $\mu_{\delta}$ = -1.495$\pm$0.242 mas yr$^{-1}$. The \textit{Gaia} DR2 results for the YSO candidates from L1147/1158 (blue), L1172/1174 (red), and L1177 (green) are shown in Fig. \ref{fig:pm_dist_yso_cepheus}. Twenty-one sources are found to form a tight group, while three sources belonging to the L1147/1158 region seem to be clear outliers. Considering that all the sources that form the tighter group fall within six times the MAD values, we used this as the criterion to select additional co-moving sources from the regions.

In order to identify the co-moving sources around BD+68$\degr$1118 and PV Cep, we first obtained the proper motions and distances of all the sources that fall within a circular region of radius 3.5$\degr$ centered on HD 200775. The choice of the search region was made based on the fact that all 94 YSO candidates identified in the vicinities of L1147/1158, L1172/1174, and L1177 fall within this circular region. While a uniform distribution of co-moving sources in the search area could imply that they are just a chance projection of field stars, a clustered distribution nearer to any of the existing star forming regions would imply that they are likely to be members of the respective star forming regions. In Paper I, we obtained a group of young co-moving sources around HD 200775 in L1172/1174. Here, we investigate if there are additional co-moving sources that are clustered around L1147/1158 and L1177. Our search provides a total of 62,637 sources within the circular area of radius 3.5$\degr$. To select the sources, we again used the same three criteria that we used for selecting the \textit{Gaia} DR2 counterparts for the known YSO candidates (m/$\sigma_{m}\geqslant$3, distance $\leqslant$1 kpc, RUWE$\leqslant$1.5). Of the 62,637 sources, 74 are found within the ellipses defined by the six times the MAD values of proper motions and distance and drawn at the median values obtained for the known YSO candidates.

\begin{figure}
    \centering
    \includegraphics[height=6.5cm, width=8.4cm]{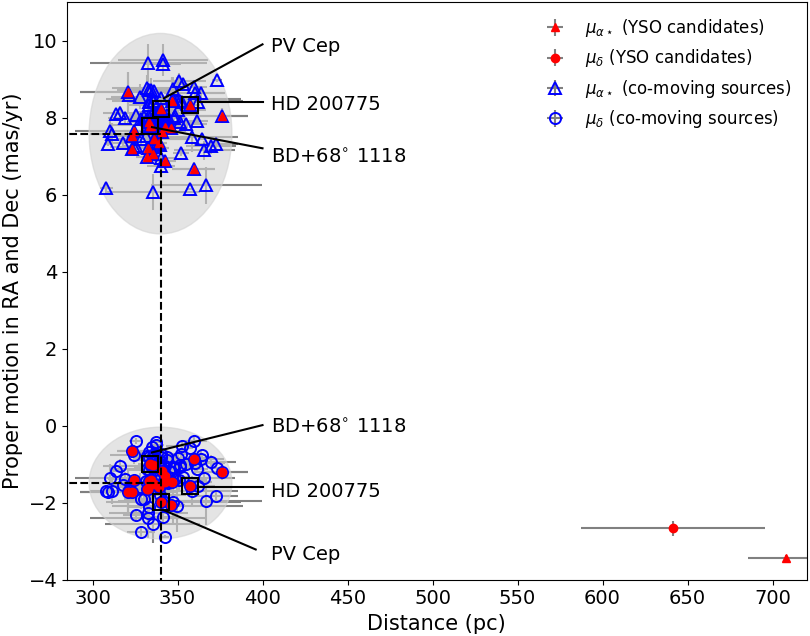}
    \caption{Proper motion vs. distance plot for the known YSO candidates and the newly identified co-moving sources. The filled triangles and circles in red represent the $d-\mu_{\alpha\star}$ and $d-\mu_{\delta}$ values, respectively, of the known YSO candidates in L1147/1158, L1172/1174, and L1177. The open blue triangles and circles are the $d-\mu_{\alpha\star}$ and $d-\mu_{\delta}$ values of the co-moving sources, respectively. The locations of HD 200775, BD+68$\degr$1118, and PV Cep are identified by square boxes. The gray ellipses represent the boundary of the proper motion values and the distance ranges used to identify the new co-moving sources. The dashed lines show the median values of $d$, $\mu_{\alpha\star}$, and $\mu_{\delta}$.}
    \label{fig:pm_dist_yso_comove_cepheus}
\end{figure}
\begin{figure}
    \centering
    \includegraphics[height=6.5cm, width=8.4cm]{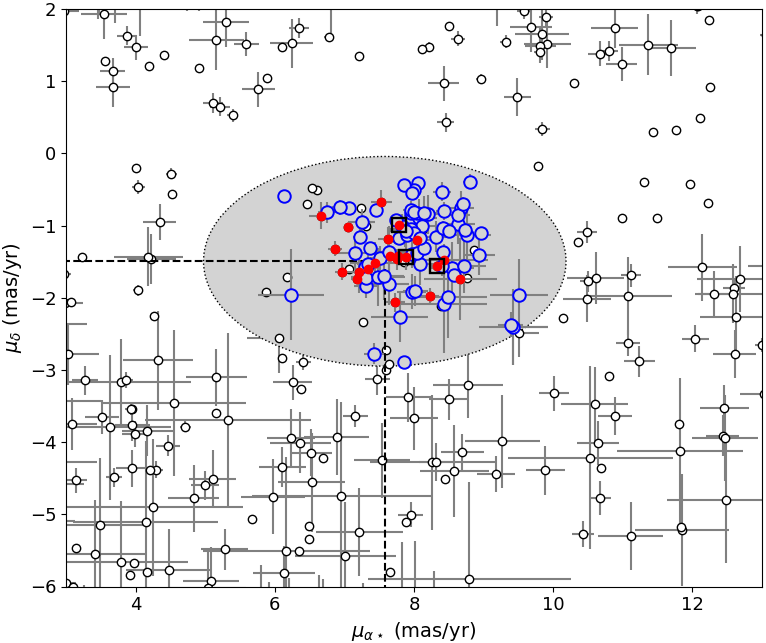}
    \caption{$\mu_{\delta}$ vs. $\mu_{\alpha\star}$ plot for the known YSO candidates and for the co-moving sources. Filled red circles represent proper motion values of known YSO candidates in L1147/1158, L1172/1174, and L1177. Open black squares mark PV Cep, HD 200775, and BD+68$\degr$1118. The open blue circles represent the same for the co-moving sources. The gray ellipse shows the boundary of the proper motion values considered when selecting the co-moving sources. The open black circles represent sources not satisfying the six times the MAD conditions in distance and proper motion values. The dashed lines show the median values of $\mu_{\alpha\star}$ and $\mu_{\delta}$.}
    \label{fig:pm_radec_yso_comove_cepheus}
\end{figure}
\begin{figure}
    \centering
    \includegraphics[height=7.6cm, width=9cm]{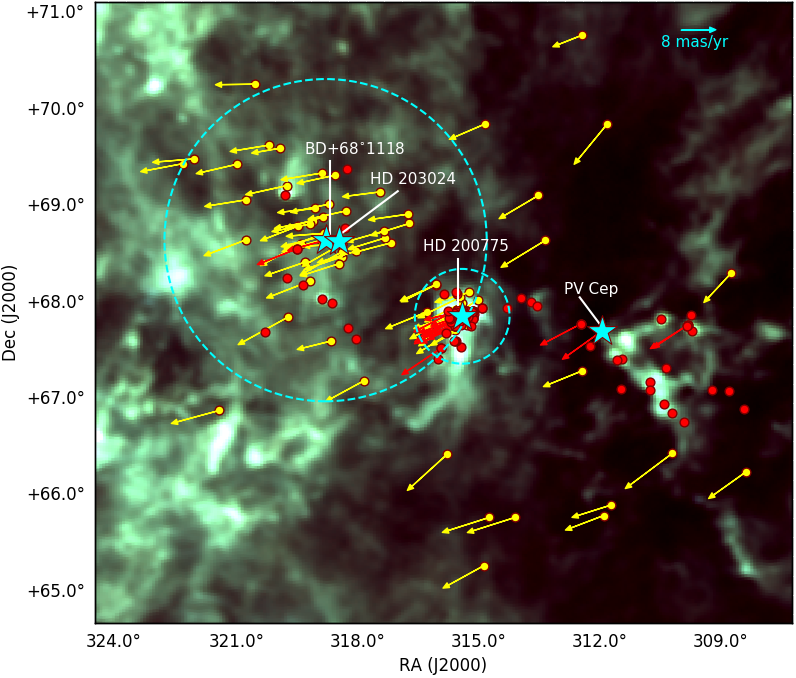}
    \caption{Proper motion plot for the YSO candidates (red arrows) and co-moving sources (yellow arrows), overplotted on the color-composite image of \textit{Planck} 353 GHz (red), 545 GHz (green), and 857 GHz (blue) images. The YSO candidates without detection in \textit{Gaia} DR2 are represented by filled red circles. HD 200775, HD 203024, BD+68$\degr$1118, and PV Cep are indicated by cyan star symbols. A vector of proper motion 8 mas yr$^{-1}$ is shown as a reference.}
    \label{fig:pm_yso_comove_cepheus}
\end{figure}

The proper motion and distance values of the 74 co-moving sources and the previously known YSO candidates are shown in Fig. \ref{fig:pm_dist_yso_comove_cepheus}. As in the case of the known YSO candidates, a similar clustering of a number of sources is apparent. The $\mu_{\alpha\star}$-$\mu_{\delta}$ values of both the co-moving and the YSO candidates are shown in Fig. \ref{fig:pm_radec_yso_comove_cepheus}. A clustering of sources is clearly noticeable when compared with the sources that fall outside the ellipses in Fig. \ref{fig:pm_radec_yso_comove_cepheus}. Some of the sources that fall inside the ellipse in Fig. \ref{fig:pm_radec_yso_comove_cepheus} but were not selected are those that do not satisfy the distance criterion. They fall just outside the ellipses drawn in Fig. \ref{fig:pm_dist_yso_comove_cepheus}. The distribution of the known YSO candidates (red circles) and the 74 co-moving sources (yellow circles) are shown in the color-composite image made using \textit{Planck} 353, 545, and 857 GHz images in Fig. \ref{fig:pm_yso_comove_cepheus}. The distribution of the co-moving sources is not uniform over the entire search area but rather concentrated mainly toward HD 200775, BD+68$\degr$1118, and HD 203024. 

\subsection{Properties of the co-moving sources}

\subsubsection{\textit{Gaia} DR2 color-magnitude diagram}\label{sec:gaia_cmd}

\begin{figure}
    \centering
    \includegraphics[height=7.2cm, width=9cm]{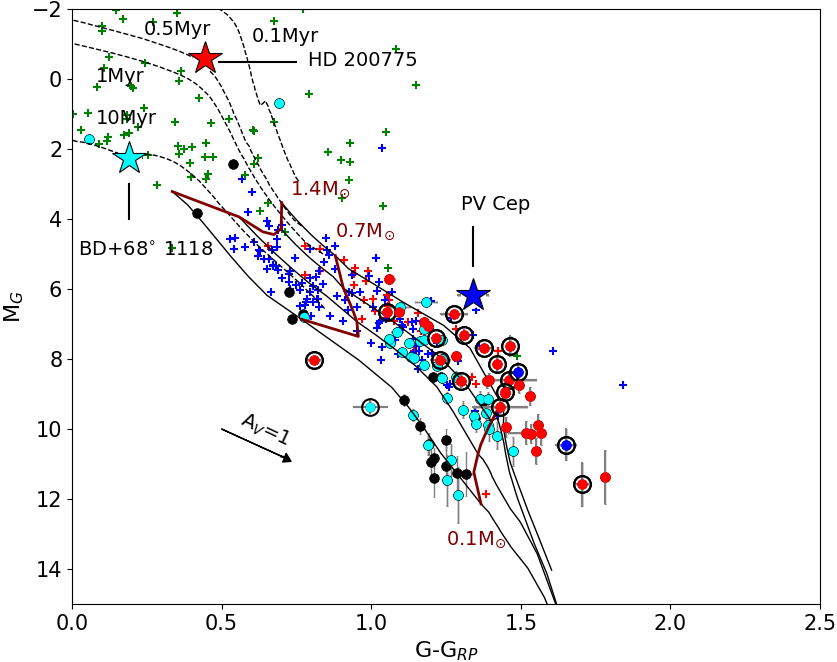}
    \caption{M$_{G}$ vs. ($G$-$G_{\mathrm{RP}}$) color-magnitude plot of the YSO candidates and the co-moving sources. The dashed lines indicate the isochrones from PARSEC models \citep{2017ApJ...835...77M} for 0.1, 0.5, 1, and 10 Myr. The solid curves represent the same from CIFIST models \citep{2015A&A...577A..42B} for 1, 3, 10, and 60 Myr. Filled blue, red, and cyan circles are the YSO candidates that have reliable \textit{Gaia} DR2 data (with open black circles) or co-moving sources (without open black circles) toward L1147/1158 (PV Cep), L1172/1174 (HD 200775), and L1177 (BD+68$\degr$1118), respectively. The co-moving sources not associated with any of the cloud complexes are presented using filled black circles. The green, red, and blue plus symbols indicate the HAeBe stars, CTTSs, and WTTSs. The arrow represents an extinction of 1 magnitude. 
    }
    \label{fig:gaia_cmd_yso_comov_cepheus}
\end{figure}

In Fig. \ref{fig:gaia_cmd_yso_comov_cepheus} we show the M$_{G}$ versus ($G$-$G_{\mathrm{RP}}$) CMD for the known YSO candidates and the newly identified co-moving sources found from this study. The $G$ and $G_{\mathrm{RP}}$ magnitudes for the 21 YSO candidates and the 74 co-moving sources are obtained from the \textit{Gaia} DR2 database. A similar plot was presented and discussed in Paper I (Fig. 13). The PMS isochrones corresponding to 0.1, 0.5, 1, 3, 10, and 60 Myr are also shown in Fig. \ref{fig:gaia_cmd_yso_comov_cepheus}. We used two grids of models, the CIFIST (Cosmological Impact of the First STars) 2011\_2015\footnote{\url{phoenix.ens-lyon.fr/Grids/BT-Settl/ CIFIST2011\_ 2015/ISOCHRONES/}} models for low-mass stars \citep[thick curves in black;][]{2015A&A...577A..42B} and the Padova tracks PARSEC (PAdova and TRieste Stellar Evolution Code) 3.3\footnote{\url{stev.oapd.inaf.it/cmd}} for the higher-mass stars \citep[dashed curves in black;][]{2017ApJ...835...77M}. The HAeBe stars, CTTSs, and WTTSs were obtained from \cite{1994A&AS..104..315T} and \cite{2010ApJ...724..835W}, respectively. Out of the 21 YSO candidates that have reliable detection in \textit{Gaia} DR2, extinction values for only 12 of them are available in \cite{2009ApJS..185..451K}. As almost half of the sources show a lack of extinction values, we made an M$_{G}$ versus ($G$-$G_{\mathrm{RP}}$) CMD without extinction correction. In Fig. \ref{fig:gaia_cmd_yso_comov_cepheus}, HD 200775, BD+68$\degr$1118, and PV Cep are identified using red, cyan, and blue star symbols, respectively. Based on its location, we inferred an age of $\sim$10 Myr for BD+68$\degr$1118, which is consistent with the 7 Myr age estimated by \cite{2009ApJS..185..451K}. We determined an age of $\sim$0.5 Myr for HD 200775, which is in agreement with the age estimates available in the literature \citep{2008MNRAS.385..391A, 2018A&A...620A.128V, 2019AJ....157..159A}. The position of PV Cep in Fig. \ref{fig:gaia_cmd_yso_comov_cepheus} seems to imply that the star is in a very early stage of evolution \citep[0.1 Myr;][]{1998A&A...334..253F}. 

Assuming that the YSO candidates and the co-moving sources are all initially formed very close to BD+68$\degr$1118, HD 200775, or PV Cep as an association or a group, we estimated the extent to which the sources could drift away due to their random velocity dispersion. Using a typical velocity dispersion of $\sim$1 kms$^{-1}$ \citep[e.g., ][]{1993AJ....105.1927G, 2015ApJ...799..136F}, the age of the intermediate-mass star, and a distance of 340 pc, the estimated values were found to be $\sim$1.7$\degr$ ($\sim$10 pc), 0.5$\degr$ ($\sim$3 pc), and 0.02$\degr$ ($\sim$0.1 pc) for BD+68$\degr$1118, HD 200775, and PV Cep, respectively (indicated by the dashed cyan circles in Fig. \ref{fig:pm_yso_comove_cepheus} for each star). Though the location of HD 200775 in Fig. \ref{fig:gaia_cmd_yso_comov_cepheus} suggests that it has an age of $\sim$0.5 Myr, the positions of the other known YSO candidates associated with the region (identified in Fig. \ref{fig:gaia_cmd_yso_comov_cepheus} using filled red circles with a black halo) suggest a range in age from $\sim$1$-$3 Myr (also see Paper I). This is consistent with the median age of the YSO candidates ($\sim$1.6 Myr) obtained by \cite{2009ApJS..185..451K}. Thus, in the case of HD 200775, we used an age of 3 Myr in our calculation.

A total of 39 co-moving sources are found within the circular region of radius 1.7$\degr$ centered on BD+68$\degr$1118, and 17 co-moving sources are found within the circular region of radius $0.5\degr$ around HD 200775. Even though a number of sources close to HD 200775 also fall within the 1.7$\degr$ circle drawn at BD+68$\degr$1118, we assigned them to the HD 200775 group because of their proximity to the star. No co-moving sources are found close to PV Cep. The \textit{Gaia} DR2 properties of the 39 co-moving sources around BD+68$\degr$1118 are presented in Table \ref{tab:YSO_new} in the appendix. The 17 co-moving sources around HD 200775 were already published in Table 4 of Paper I, except \#c1, \#c2, and \#c9; these 17 sources are therefore not presented here. The 18 co-moving sources that are not associated with any of these three regions are listed in Table \ref{tab:YSO_new} in the appendix. As presented in Paper I, the co-moving sources belonging to HD 200775 show a well-defined sequence roughly following the 1$-$3 Myr isochrones, similar to the distribution of the known YSO candidates. In Paper I, we found that the spectroscopy of three co-moving sources belonging to the HD 200775 region shows H$\alpha$ in emission; this is considered to be an indicator of their youth as the line is formed mainly due to the accretion process and stellar magnetic activity \citep{1994AJ....108.1056E, 1998AJ....116..455M}. It is evident that the co-moving sources surrounding BD+68$\degr$1118 show an age of $\sim$10 Myr, which is consistent with the age of BD+68$\degr$1118 itself. 

\subsubsection{The $(g-r)$ versus $(r-i)$ color-color diagram}\label{sec:panstarrs}

\begin{figure}
    \centering
    \includegraphics[height=7cm, width=9cm]{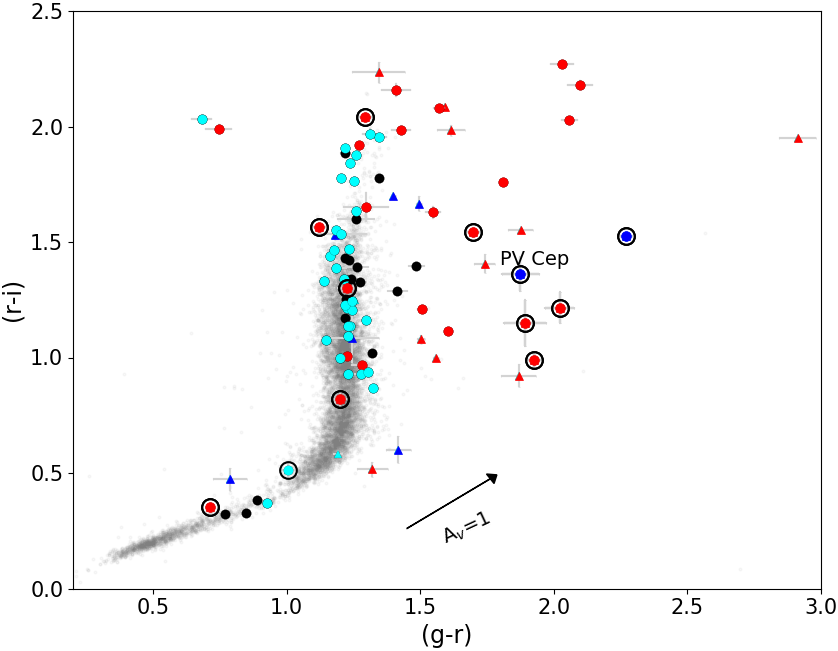}
    \caption{ (r-i) versus (g-r) CC diagram for the known YSO candidates and the newly identified co-moving sources. Filled cyan, red, and blue circles represent co-moving sources identified toward BD+68$\degr$1118, HD 200775, and PV Cep, respectively. Filled cyan, red, and blue circles with open black circles represent known YSO candidates identified toward BD+68$\degr$1118, HD 200775, and PV Cep, respectively. Filled cyan, red, and blue triangles represent YSO candidates not detected or without reliable \textit{Gaia} DR2 data identified toward BD+68$\degr$1118, HD 200775, and PV Cep, respectively. The co-moving sources not associated with any of the three regions are presented using filled black circles. The arrow represents the reddening vector corresponding to an A$_{V}$ of 1 magnitude.}
    \label{fig:gri_CC_yso_cepheus}
\end{figure}

In Fig. \ref{fig:gri_CC_yso_cepheus} we present a $(g-r)$ versus $(r-i)$ CC diagram for the sources studied here based on the data obtained from the Pan-STARRS catalog \citep{2016arXiv161205560C}, using a search radius of $1^{\prime\prime}$ for individual sources. The gray dots represent the loci of main sequence stars selected from a region around the star 10 Lac that have spectral types in the range of A0 to M7. The spectral types of the sources were obtained from the Simbad database. We selected this region because the E(B-V) values estimated for 10 Lac \citep[$\sim$ 0.08; ][]{2002BaltA..11....1W, 2003AN....324..219W, 2018A&A...613A...9M} suggest that the direction toward it presents a region of low extinction at least up to the distance of the star \citep[$\sim$ 450 pc; ][]{2020yCat.1350....0G}. Of the 94 YSO candidates found in the region, we obtained $g$, $r$, and $i$ data for 30 sources, 12 of which  also have \textit{Gaia} DR2 data. Out of the 74 co-moving sources found in the region, we obtained $g$, $r$, and $i$ data for 65 sources. 

Reddened and un-reddened M-dwarfs occupy a distinct locus in the $(g-r)$ versus $(r-i)$ CC diagram compared to the rest of the early main sequence stars and the giants, allowing us to distinguish them clearly. From Fig. \ref{fig:gri_CC_yso_cepheus}, it is apparent that a majority of the sources identified as known YSO candidates and co-moving sources are M-dwarfs. Of these, the co-moving sources around BD+68$\degr$1118 and those not belonging to any of the three regions follow un-reddened M-dwarf loci, suggesting a negligible foreground extinction. Some of the YSO candidates may have a contribution from circumstellar material apart from the foreground extinction. Sources belonging to the HD 200775 and PV Cep regions are the ones that show relatively high extinction. 

\subsubsection{Near-IR and mid-IR properties}

\begin{figure}
    \centering
    \includegraphics[height=7cm, width=9cm]{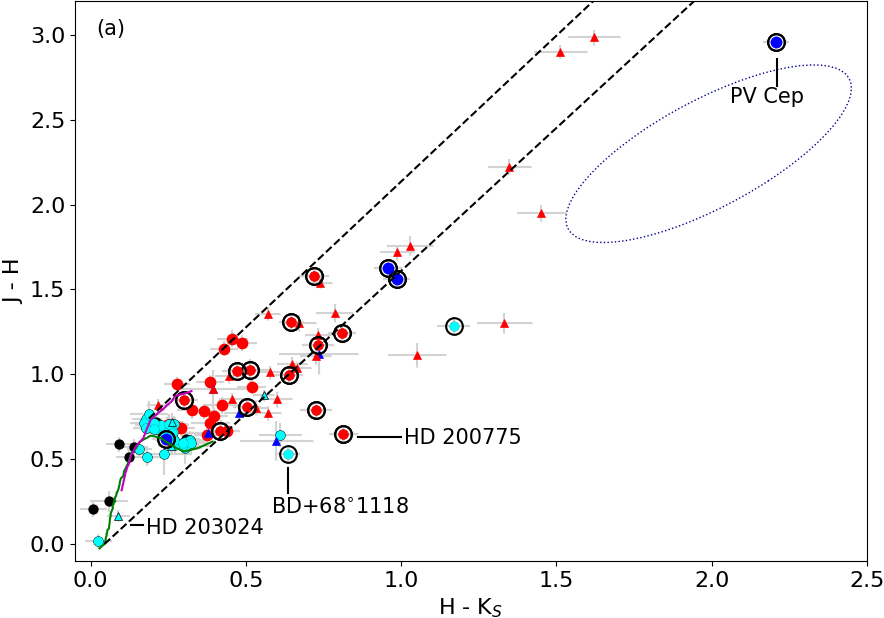}
    \includegraphics[height=7cm, width=9cm]{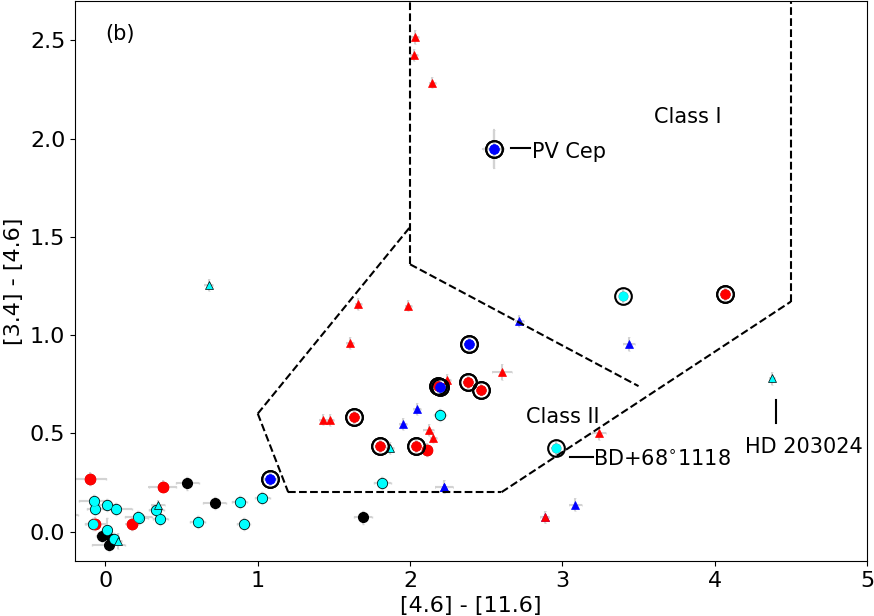}
    \caption{Near- and mid-IR CC diagrams for the known YSO candidates and the newly identified co-moving sources. Symbols are the same as presented in Fig. \ref{fig:gri_CC_yso_cepheus}. \textbf{(a)} (\textit{J$-$H}) vs. (\textit{H$-$K$_{S}$}) CC diagram of the sources. The solid curves in green and magenta represent the loci of the un-reddened main sequence stars and the giants \citep{2010A&A...509A..44M}, respectively. The ellipse represents the range of variability of PV Cep, based on the time series observations by \cite{2011ApJ...732...69L}. \textbf{(b)} \textit{WISE} CC diagram of the sources. The dashed lines separate the regions occupied by the Class I and Class II sources \citep{2014ApJ...791..131K}.}
    \label{fig:jhk_wise_CC_yso_cepheus}
\end{figure}

We investigated the near- and mid-IR properties of the newly found co-moving sources by obtaining their 2MASS and \textit{WISE} magnitudes from the \cite{2003yCat.2246....0C} and \cite{2012yCat.2311....0C} catalogs, respectively. The region surrounding BD+68$\degr$1118 was not observed by the \textit{Spitzer} satellite. Out of the 74 co-moving sources, we found 2MASS counterparts for 65 of them within a search radius of $1\arcsec$. Of the 21 known YSO candidates with reliable \textit{Gaia} DR2 data, we found 2MASS data for 18 of them. The known YSO candidates with reliable \textit{Gaia} DR2 data and the co-moving sources are shown in the (\textit{J$-$H}) versus (\textit{H$-$K$_{S}$}) CC diagram in Fig. \ref{fig:jhk_wise_CC_yso_cepheus} (a). While a majority of the known YSO candidates show a relatively high amount of extinction and near-IR excess, the newly identified co-moving sources exhibit relatively low extinction and show a small or negligible amount of near-IR excess emission. Based on the near-IR photometric study of PV Cep, \cite{2011ApJ...732...69L} found a significant variability at different epochs, the range of which is shown by an ellipse. The star shows extremely high near-IR excess and extinction, which is consistent with its very young age ($\lesssim$ 1 Myr). The sources HD 200775 and BD+68$\degr$1118 occupy the region normally occupied by HAeBe stars. It is intriguing to note that, despite being a $\sim$10 Myr star, BD+68$\degr$1118 still shows significant near-IR excess. The co-moving sources around BD+68$\degr$1118, on the other hand, show little to no extinction or near-IR excess emission. A significant number of co-moving sources around HD 200775 show evidence of some amount of extinction but negligible near-IR excess emission. A majority of the co-moving sources found in the vicinity of BD+68$\degr$1118 are located along the loci occupied by the M-dwarfs, which is consistent with the deductions in Fig. \ref{fig:gri_CC_yso_cepheus}.

\begin{figure}
    \centering
    \includegraphics[height=7cm, width=9cm]{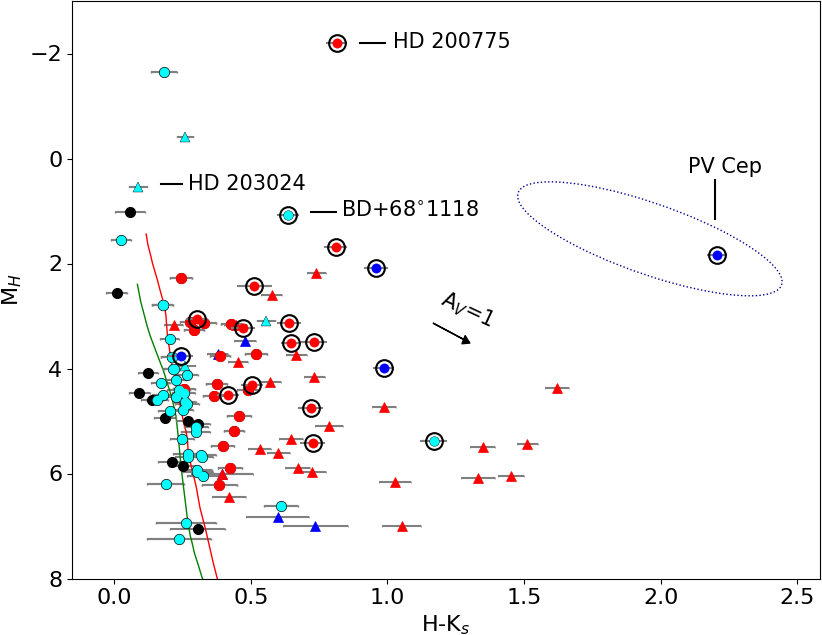}
    \caption{M$_{H}$ vs. \textit{H$-$K$_{S}$} CMD for the known YSO candidates and the newly identified co-moving sources. Symbols are the same as presented in Fig. \ref{fig:gri_CC_yso_cepheus}. The arrow represents an extinction of 1 magnitude. The ellipse has the same meaning as in Fig. \ref{fig:jhk_wise_CC_yso_cepheus} (a).}
    \label{fig:hhk_CM_yso_cepheus}
\end{figure}

The \textit{WISE} magnitudes of the sources were obtained from the catalog provided by \cite{2012yCat.2311....0C}, with a search radius of 3$^{\prime\prime}$. We selected only the \textit{W1}, \textit{W2,} and \textit{W3} bands since the majority of magnitudes given in the \textit{W4} band are only an upper limit. We verified the detection of the sources in each of the bands by inspecting them visually in the \textit{WISE} images.\ We did this because we noticed that for a significant number of sources, though the magnitudes in the \textit{W3} and \textit{W4} $\mu$m bands are provided in the \textit{WISE} catalog, no detection was found in the corresponding images. The results are shown in Fig. \ref{fig:jhk_wise_CC_yso_cepheus} (b) in the [3.4]$-$[4.6] versus [4.6]$-$[11.6] CC diagram. Of the 74 co-moving sources, we found a counterpart for 29 in the \textit{WISE} database. Out of the 21 YSO candidates with reliable \textit{Gaia} DR2 data, we obtained \textit{WISE} magnitudes for 13. Of the remaining known YSO candidates devoid of reliable data or without \textit{Gaia} DR2 detections, we obtained \textit{WISE} magnitudes for 24 sources. The distribution of the known YSO candidates and the co-moving sources are more distinguishable in Fig. \ref{fig:jhk_wise_CC_yso_cepheus} (b). Evidently, the known YSO candidates fall in a region generally populated by Class I and Class II objects. A majority of the newly identified co-moving sources fall in a region generally occupied by the Class III sources, consistent with our deductions in Figs. \ref{fig:gri_CC_yso_cepheus} and  \ref{fig:jhk_wise_CC_yso_cepheus} (a). 

In Fig. \ref{fig:hhk_CM_yso_cepheus} we present an M$_{H}$ versus (\textit{H$-$K$_{S}$}) CMD for the previously known YSO candidates and the newly found co-moving sources. The PMS isochrones of 1 and 10 Myr are taken from the CIFIST models and are shown as red and green curves, respectively. The extent of variability observed for PV Cep \citep{2011ApJ...732...69L} is again shown here using an ellipse. Evidently, the sources located in the vicinity of HD 200775 show higher values of (\textit{H$-$K$_{S}$}) colors compared to those of the sources identified toward the BD+68$\degr$1118 association. The higher (\textit{H$-$K$_{S}$}) colors of the sources toward HD 200775 could be due to the contribution of a substantial amount of circumstellar and interstellar material present along the line of sight, which is consistent with their relatively young ages.

\subsection{Spatial distribution of sources surrounding HD 200775 and BD+68$\degr$1118}

HD 203024, which is located at an angular distance of $\sim$8.5$\arcmin$ (0.9 pc) west of BD+68$\degr$1118, is identified as a spectroscopic binary \citep{2013MNRAS.429.1001A}. The effective temperatures and masses of the components are found to be 9250 K and 6500 K and 2.8 M$_{\odot}$ and 1.6 M$_{\odot}$, respectively. The ages of the components are estimated as 9.3 Myr and 2.7 Myr, respectively \citep{2013MNRAS.429.1001A}. The binary nature of HD 203024 could be the reason why the parallax and proper motion solutions are not listed in \textit{Gaia} DR2. \citet{1997AstL...23...97M} and \citet{2000MNRAS.319..777K} detected H$\alpha$ in emission. The assigned spectral type of the star ranges from B8.5V to A5V \citep{1997AstL...23...97M, 2001A&A...378..116M, 2013MNRAS.429.1001A}. The star was classified as a HAeBe candidate by \citet{2000MNRAS.319..777K}, and the presence of a debris disk \citep{2016A&A...594A..59R} surrounding the star suggests that it is at an advanced stage of evolution. The proper motion values derived by \cite{2011MNRAS.416..403F} for HD 203024 are found to be $\mu_{\alpha\star}$ = 9.89 mas yr$^{-1}$ and $\mu_{\delta}$ = -2.10 mas yr$^{-1}$, which are in good agreement with those of BD+68$\degr$1118, HD 200775, and PV Cep. No additional stellar association, lying within similar distance and proper motion ranges, was found, even within twice the search radius. Thus, BD+68$\degr$1118, HD 200775, and PV Cep together form an isolated intermediate-mass stellar association that is connected both physically and kinematically, but they are distributed over a wide area in the sky and show a $\sim$1$-$10 Myr age spread.

\begin{figure}
    \centering
    \includegraphics[height=7cm, width=8.7cm]{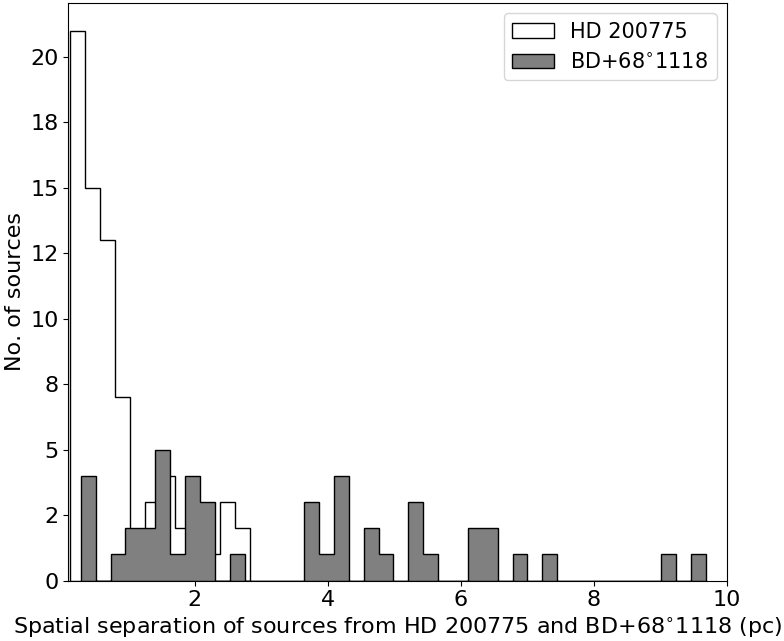}
    \caption{Histograms of the spatial distribution of the known YSO candidates and the newly identified co-moving sources with respect to HD 200775 (white) and BD+68$\degr$1118 (gray).}
    \label{fig:HD200775_radial_profile_cepheus}
\end{figure}

Although BD+68$\degr$1118, HD 200775, and PV Cep are at similar distances and share similar proper motions, they show some stark differences. The BD+68$\degr$1118 region lacks molecular cloud material within the radial velocity range of 2.7-2.9 km~s$^{-1}$ associated with it (except L1177), in contrast to the regions surrounding HD 200775 and PV Cep \citep{1997ApJS..110...21Y}. The known YSO candidates and the newly identified co-moving sources surrounding BD+68$\degr$1118 are distributed over a larger area in the sky in comparison with those surrounding HD 200775, as shown in Fig. \ref{fig:HD200775_radial_profile_cepheus}. While all the sources surrounding HD 200775 are distributed within a spatial distance of $\sim$3 pc from it, those surrounding BD+68$\degr$1118 are distributed over a distance of up to $\sim$10 pc. The sources around BD+68$\degr$1118 lack near- and mid-IR excess emission, while a significant number of sources around HD 200775 show excess emission. This is consistent with the findings that the sources around BD+68$\degr$1118 are more evolved, showing an age of $\sim$3$-$10 Myr, while those around HD 200775 are comparatively young ($\sim$1$-$3 Myr).

\begin{figure}
    \centering
    \includegraphics[height=7cm, width=8cm]{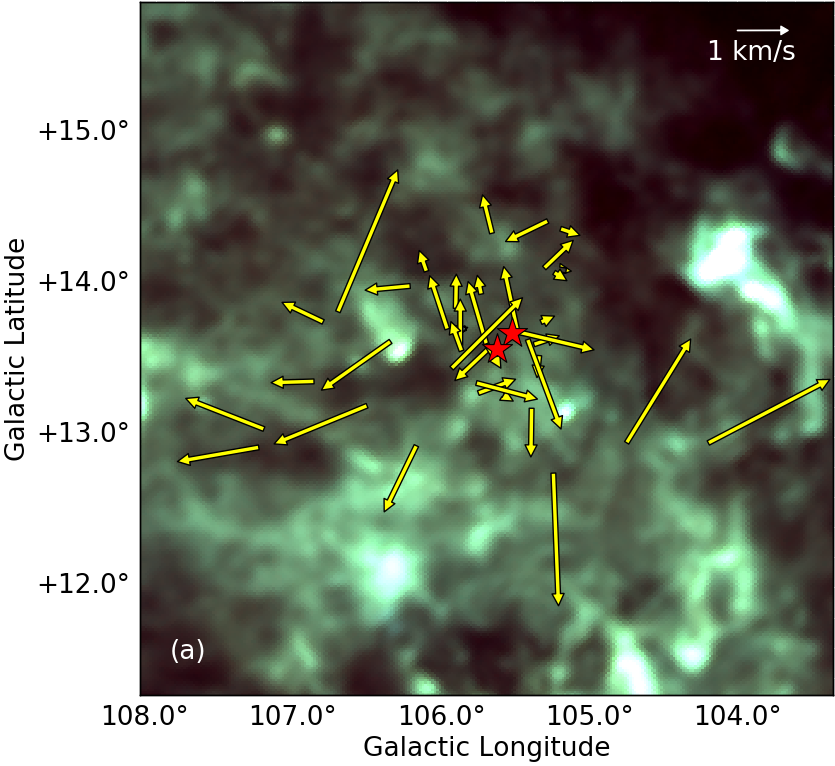}
    \includegraphics[height=7cm, width=8cm]{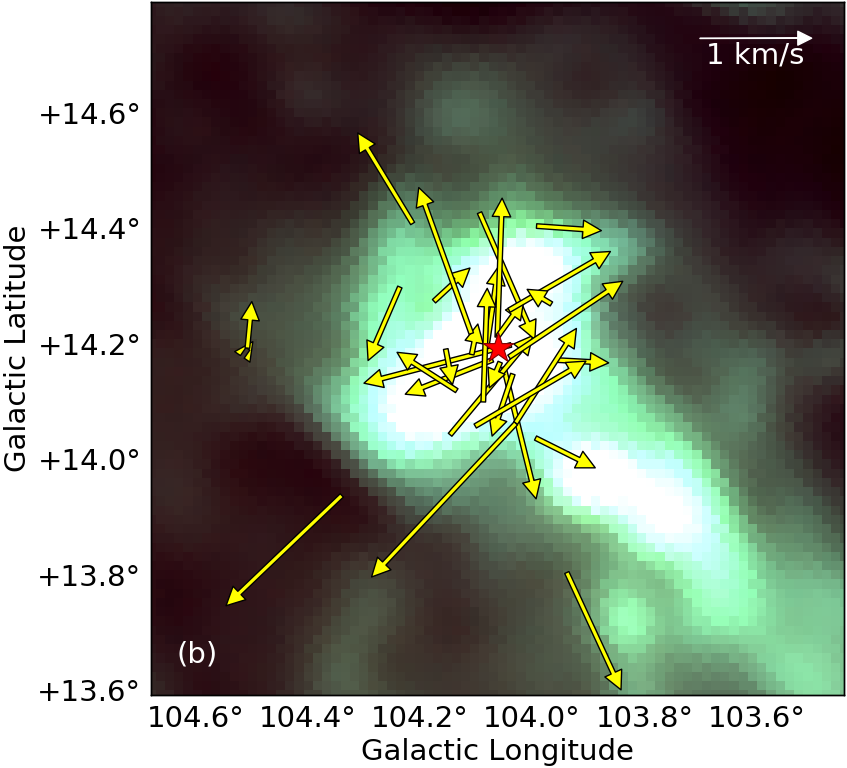}
    \caption{Proper motion vectors of YSO candidates and co-moving sources relative to the system centers shown using yellow arrows overplotted on the color-composite image of \textit{Planck} 353 GHz (red), 545 GHz (green), and 857 GHz (blue) images. (a) Relative proper motion of the sources around BD+68$\degr$1118. The positions of BD+68$\degr$1118 and HD 203024 are presented using red star symbols. (b) Same as (a), but for the HD 200775 system (marked as a red star).}
    \label{fig:pm_yso_cepheus}
\end{figure}


\subsection{Kinematic properties of the co-moving sources}

Star formation is an inefficient process, and hence the conversion of gas and dust to form stars is never complete. The unconsumed gas and dust are believed to get expelled out of a star cluster through stellar feedback in the form of photoionization radiation, stellar outflows, winds from massive stars, and the onset of the first supernova explosion \citep{1953PASP...65..205Z, 1997MNRAS.284..785G, 2000ApJ...545..364M, 2006epbm.book.....E, 2013A&A...559A..38P}. This impulsive gas expulsion could lead the stellar components to acquire higher velocities and become unbound, and the whole system might eventually dissolve into the field, feeding the field star population \citep[e.g.,][]{2011MNRAS.415.3439D, 2013A&A...559A..38P, 2013MNRAS.436.3727D}. In order to investigate the motion of the stellar associations of BD+68$\degr$1118 and HD 200775, we considered the proper motions of the member stars, which show the movement projected in the plane of the sky as a whole. However, the proper motion values of each member of the group subtracted from the mean proper motion of the whole group determines their projected internal motion \citep{1997MmSAI..68..833J}. In order to analyze the internal motion of the sources surrounding BD+68$\degr$1118 and HD 200775, we studied their motion relative to the system center.

Panels (a) and (b) of Fig. \ref{fig:pm_yso_cepheus} represent the relative proper motion (yellow vectors) of the sources distributed around BD+68$\degr$1118 and HD 200775, respectively, in Galactic coordinates. The members of the BD+68$\degr$1118  stellar association  move outward in all directions, while a majority of the sources surrounding HD 200775 tend to move in the northwestern and southeastern directions; these are also the directions of outflow for HD 200775 \citep{1998A&A...339..575F}, which is currently inactive. The velocity components $v_{l}$ and $v_{b}$ along the Galactic longitude and latitude were computed using $v_{l} = 4.74d\mu_{l}$ and $v_{b} = 4.74d\mu_{b}$, respectively. The factor $4.74\times d$ (kpc) transforms the unit of proper motion from mas yr$^{-1}$ into km s$^{-1}$. The possible compression and expansion of the  BD+68$\degr$1118 and HD 200775 associations were calculated using the \textit{Gaia} DR2 proper motions of the surrounding YSO candidates and co-moving sources. The parameters of compression and expansion were calculated following the methods provided by \cite{2017MNRAS.472.3887M}:\\
\begin{equation}
    v_{l} = v_{l0} + p_{l}~d sin(l-l_{0})
,\end{equation} 
\begin{equation}
    v_{b} = v_{b0} + p_{b}~d sin(b-b_{0})
,\end{equation}
\noindent
where: $v_{l0}$ and $v_{b0}$ are the average velocities of the sources in the $l$ and $b$ directions, respectively; $l_{0}$ and $b_{0}$ are the coordinates of the center of the association; $p_{l}$ or $p_{b}$ are the parameters that indicate the expansion and compression along the $l$ and $b$ directions (positive or negative values signify an expansion or compression of the system, respectively); and $d$ is the distance to the association. Using Eqs. (1) and (2), we estimated mean $p_{l}=175\pm21$ and $p_{b}=80\pm40$ km s$^{-1}$ kpc$^{-1}$ for the BD+68$\degr$1118 association and $p_{l}=270\pm170$ and $p_{b}=360\pm160$ km s$^{-1}$ kpc$^{-1}$ for HD 200775 association. These values signify that both the systems are expanding. The errors are estimated by propagating uncertainties in the parameters. The observed velocity of expansion in the $l$ and $b$ directions are determined using the following equations, provided by \cite{2017MNRAS.472.3887M}:\\
\begin{equation}
    u_{l} = p_{l}a
,\end{equation} 
\begin{equation}
    u_{b} = p_{b}a
,\end{equation}
where $a$ is the radius of the association in kpc. We estimated $a$ as the radius containing 99.7 per cent of the associated members. Therefore, the computed radii for the BD+68$\degr$1118 and HD 200775 associations are $\sim$8 and $\sim$2 pc, respectively. The estimated $u_{l}$ and $u_{b}$ values for the association of BD+68$\degr$1118 are $1.4\pm0.2$ and $0.7\pm0.3$ km s$^{-1}$, and for HD 200775 they are $0.5\pm0.3$ and $0.7\pm0.3$ km s$^{-1}$, respectively. The locations of the two sparse clusters associated with BD+68$\degr$1118 and HD 200775 on an age-radius plane \citep{2009A&A...498L..37P, 2013A&A...559A..38P} suggest that they form the lower end of the correlation found for the loose clusters. We also estimated the tangential components of the expansion velocities ($v_{t}$) of the BD+68$\degr$1118 system as $1.5\pm0.2$, and for HD 200775 system as $0.9\pm0.3$ km s$^{-1}$, following $v_{t}=\sqrt{u_{l}^{2}+u_{b}^{2}}$. 

L1147/L1158, L1172/1174, and L1177 are located outside of the Cepheus Flare Shell (CFS) but near the periphery of Loop III \citep{2007IAUS..237..119K}. Located at $\sim$300 pc, these huge, nearly concentric shells are considered to be effects of multiple supernova explosions, indicating the presence of OB stars a few million years ago \citep{2006MNRAS.369..867O, 2007IAUS..237..119K}. The spatial distribution of YSO candidates in the star forming centers toward the CFS is located close to the edges of the molecular clouds, indicating star formation possibly induced by an expansion of the shell \citep{1998ApJS..115...59K, 2007IAUS..237..119K}. In our study, we found the age of the YSO candidates to be 10 Myr toward L1177 (the BD+68$\degr$1118 stellar association), which is located close to the center of Loop III and the CFS. The sources with ages in the $\sim$1$-$3 Myr range are distributed in L1172/1174 (HD 200775 stellar association), which is located in an intermediate area, and the extreme young source PV Cep ($\sim$0.1 Myr), embedded in L1147/1158, lies at the outer edge of Loop III and the CFS. Therefore, the spatial distribution of the stellar age revealed in our work further confirms that triggered star formation is currently occurring in the CFS. 

\section{Summary and conclusion} \label{sec:sum_con}

We studied the kinematics of the entire region containing four intermediate-mass young sources, namely BD+68$\degr$1118, HD 203024, HD 200775, and PV Cep. Using \textit{Gaia} DR2 distance and proper motion measurements, we identified new co-moving sources surrounding BD+68$\degr$1118 and HD 203024. Combining the new co-moving sources identified toward HD 200775 (Paper I) with the new co-moving sources identified in this work, we made an attempt to understand the star formation history of the region as a whole. Our main results are summarized below.

    \begin{itemize}
        \item Our search for co-moving sources using the \textit{Gaia} DR2 distance and proper motion values within a circular region of 3.5$\degr$ radius, which contains BD+68$\degr$1118, HD 200775, PV Cep, and the known low-mass YSO candidates, resulted in the identification of 74 sources. Of these, 39 are found to be distributed around BD+68$\degr$1118, 17 are distributed around HD 200775, and the rest are distributed over a wider region. No co-moving sources are found around the much younger PV Cep.
        \item Based on the \textit{Gaia} DR2 $G$ versus ($G-G_{\mathrm{RP}}$) CMD, near- and mid-IR CC diagrams, and near-IR CMD, the co-moving sources identified around BD+68$\degr$1118 are found to be older ($\sim$10 Myr) than those found around HD 200775 ($\sim$1$-$3 Myr). The co-moving sources are mainly M-dwarfs that show no (or negligible) near- or mid-IR excess emission.
        \item The positive values of the coefficients $p_{l}$ and $p_{b}$ for the stellar associations surrounding BD$+68\degr1118$ and HD 200775 indicate that both systems are in an expanding phase, with similar tangential velocities.
        \item The decrease in the age of the sources ($\sim$10, 3, and 0.1 Myr for the sources located toward L1177, L1172/1174, and L1147/1158, respectively) with increasing distance from the centers of the CFS and Loop III agrees with previous studies that discussed about the ongoing triggered star formation caused by external impacts.\\

More rigorous study of the newly identified co-moving sources are required to estimate their physical properties (e.g., extinction, spectral type, mass, and luminosity). It will help us to find their true characteristics and understand the triggered star formation in this region.

\end{itemize} 

\begin{acknowledgements}
    This work has made use of data from the following sources: (1) European Space Agency (ESA) mission \textit {Gaia} (\url{https://www.cosmos.esa.int/gaia}), processed by the {\it Gaia} Data Processing and Analysis Consortium (DPAC, \url{https://www.cosmos.esa.int/web/gaia/dpac/consortium}). Funding for the DPAC has been provided by national institutions, in particular the institutions participating in the {\it Gaia} Multilateral Agreement; (2) the Pan-STARRS1 Surveys (PS1) have been made possible through contributions of the Institute for Astronomy, the University of Hawaii, the Pan-STARRS Project Office, the Max-Planck Society and its participating institutes, the Max Planck Institute for Astronomy, Heidelberg and the Max Planck Institute for Extraterrestrial Physics, Garching, The Johns Hopkins University, Durham University, the University of Edinburgh, Queen's University Belfast, the Harvard-Smithsonian Center for Astrophysics, the Las Cumbres Observatory Global Telescope Network Incorporated, the National Central University of Taiwan, the Space Telescope Science Institute, the National Aeronautics and Space Administration under Grant No. NNX08AR22G issued through the Planetary Science Division of the NASA Science Mission Directorate, the National Science Foundation under Grant No. AST-1238877, the University of Maryland, and Eotvos Lorand University (ELTE); (3) the Two Micron All Sky Survey, which is a joint project of the University of Massachusetts and the Infrared Processing and Analysis Center/California Institute of Technology, funded by the National Aeronautics and Space Administration and the National Science Foundation; (4) The \textit{Wide-field Infrared Survey Explorer}, which is a joint project of the University of California, Los Angeles, and the Jet Propulsion Laboratory/California Institute of Technology, funded by the National Aeronautics and Space Administration. We also used data provided by the SkyView, which is developed with generous support from the NASA AISR and ADP programs (P.I. Thomas A. McGlynn) under the auspices of the High Energy Astrophysics Science Archive Research Center (HEASARC) at the NASA/ GSFC Astrophysics Science Division.
\end{acknowledgements}

   \bibliographystyle{aa} 
   \bibliography{ref}

\begin{appendix}
        \section{Tables}
          
\begin{table*}[ht]
    \caption{Properties of the known YSO candidates identified towards L1147/1158 and L1177 from \textit{Gaia} DR2.}
    \label{tab:YSO_gaia_cepheus}
    \renewcommand{\arraystretch}{1.3}
    \tiny
    \begin{tabular}{p{0.1cm}p{1.2cm}p{1.1cm}p{2.6cm}p{0.9cm}p{1.6cm}p{1.6cm}p{1.8cm}p{0.8cm}p{0.8cm}p{0.8cm}p{0.8cm}} 
    \hline
    \#& RA(2015.5)  & Dec(2015.5)  & Source & Distance & $\mu_{\alpha\star}$ ($\Delta\mu_{\alpha\star}$) & $\mu_{\delta}$ ($\Delta\mu_{\delta}$) & G ($\Delta$G) & RUWE & Pan- & 2MASS & \textit{WISE} \\
     & ($^{\circ}$) & ($^{\circ}$) & ID & (pc)& (mas yr$^{-1}$) & (mas yr$^{-1}$) & (mag) & &STARRS\\
     (1)&(2)&(3)&(4)&(5)&(6)&(7)&(8)&(9)&(10)&(11)&(12)\\
    \hline
    \multicolumn{11}{c}{sources associated with L1177} \\
    1 & 319.413344 & 68.919361 & 2270536045876468096 & 334$^{+2}_{-3}$ & 7.778$\pm$0.042 & -0.992$\pm$0.040 & 9.8782$\pm$0.0004 & 1.0 & - &\checkmark & \checkmark \\
    2 & 320.241149 & 68.805102 & 2222486906706076544 & 333$^{+9}_{-8}$ & 7.888$\pm$0.141 & -1.440$\pm$0.175 & 16.9815$\pm$0.0164 & 1.2 & \checkmark &\checkmark & \checkmark\\
    \hline
    \multicolumn{11}{c}{sources associated with L1147/1158} \\
    1* & 307.791609 & 67.007548 & 2246449594402553088 & 708$^{+23}_{-22}$&-3.443$\pm$0.087&-4.257$\pm$0.084&16.4659$\pm$0.0008 & 1.0 &  \checkmark &\checkmark & -\\
    2* & 308.579103 & 67.241341 & 2246818824150192768 & 824$^{+10}_{-9}$&-9.152$\pm$0.029&-6.229$\pm$0.031&14.2628$\pm$0.0003 & 1.1 &  \checkmark&\checkmark & \checkmark\\
    3* & 308.949753 & 68.048996 & 2247157198854324224 & 641$^{+59}_{-49}$&-4.395$\pm$0.225&-2.670$\pm$0.190&17.7134$\pm$0.0017 & 1.1 &  \checkmark&- & \checkmark\\
    4 & 309.048672 & 67.952608 & 2246967838039390848 & 336$^{+23}_{-20}$&7.436$\pm$0.351&-1.514$\pm$0.284&18.0815$\pm$0.0108 & 1.0 & - &\checkmark & \checkmark\\
    5 & 309.082855 & 67.942131 & 2246967876695385728 & 343$^{+14}_{-13}$&7.753$\pm$0.189&-1.469$\pm$0.143&16.0569$\pm$0.0040 & 1.2 & \checkmark &\checkmark & -\\
    6 & 311.474902 & 67.960735 & 2246924068029363840 & 341$^{+7}_{-7}$&8.228$\pm$0.126&-1.976$\pm$0.110&13.8423$\pm$0.0163 & 1.3 & \checkmark &\checkmark & \checkmark\\
    7 & 312.042682 & 68.050417 & 2270941147188904704 & 347$^{+25}_{-22}$&8.426$\pm$0.637&-1.476$\pm$0.221&15.9581$\pm$0.0019 & - &- &\checkmark & \checkmark\\
    \hline
    \end{tabular}\\
    \renewcommand{\arraystretch}{1}
    Columns 2 \& 3: 2015.5 epoch Right Ascension \& Declination of sources given by \textit{Gaia} DR2.\\
    Column 4: Source id taken from \textit{Gaia} DR2.\\
    Column 5: Distance taken from the \citet{2018AJ....156...58B} catalogue.\\
    Columns 6 \& 7: Proper motion in Right Ascension \& Declination of sources given by \textit{Gaia} DR2.\\
    Column 8: G magnitude of sources given by \textit{Gaia} DR2.\\
    Column 9: Renormalised Unit Weight Error (RUWE) of sources obtained from \textit{Gaia} DR2.\\
    $^{*}$ Sources considered as outliers in our analysis.\\
\end{table*}
\begin{table*}[ht]
	\hspace{-1.5cm}
    \caption{Properties of the co-moving sources identified towards L1147/1158 and L1177 from \textit{Gaia} DR2.}
    \label{tab:YSO_new} 
    \renewcommand{\arraystretch}{1.3}
    \tiny
    \begin{tabular}{p{0.7cm}p{1.2cm}p{1.1cm}p{2.6cm}p{0.8cm}p{1.3cm}p{1.6cm}p{1.8cm}p{0.8cm}p{0.8cm}p{0.8cm}p{0.8cm}} 
    \hline
    $\#$& RA(2015.5) & Dec(2015.5) & Source & Distance & $\mu_{\alpha\star}$ ($\Delta\mu_{\alpha\star}$) & $\mu_{\delta}$ ($\Delta\mu_{\delta}$) & G (eG) & RUWE &Pan-&2MASS & \textit{WISE}\\
     &($^{\circ}$) & ($^{\circ}$) & ID & (pc) & (mas yr$^{-1}$) & (mas yr$^{-1}$) & (mag) & & STARRS & &\\
     (1)&(2)&(3)&(4)&(5)&(6)&(7)&(8)&(9)&(10)&(11)&(12)\\
    \hline
     \multicolumn{12}{c}{sources associated with L1177} \\
     bd\_c1 &316.970365	&69.135322	& 2270674545683908224	&340$^{+13}_{-13}$ 	&8.110$\pm$0.195	&-1.011$\pm$0.204	&18.1137$\pm$0.0016	&0.9& \checkmark&\checkmark&-\\
     bd\_c2 & 317.016704 & 69.232990 &2270699666950862336 &353$^{+9}_{-9}$ & 8.401$\pm$0.127 & -0.533$\pm$0.132 & 16.2560$\pm$0.0010&1.4& \checkmark&\checkmark& \checkmark\\
     bd\_c3 &317.485774	&68.926013	&2270478145419384832	&344$^{+20}_{-18}$	&8.068$\pm$0.264	&-0.894$\pm$0.264	&17.8806$\pm$0.0015	&1.1& \checkmark&\checkmark&-\\
     bd\_c4 &317.660746	&68.976917	&2270501823576986752	&355$^{+13}_{-11}$ & 7.821$\pm$0.155	&-1.007$\pm$0.164	&17.2027$\pm$0.0013	&1.1& \checkmark&\checkmark&-\\      
     bd\_c5 &317.719747	&69.045445	&2270505396989771904	&339$^{+4}_{-3}$ &	7.741$\pm$0.053	&-0.928$\pm$0.058	&15.2079$\pm$0.0009	&1.1& \checkmark&\checkmark& \checkmark\\
     bd\_c6 &317.880961	&69.457870	&2270709356397123328	&337$^{+5}_{-5}$ &	8.003$\pm$0.070	&-0.504$\pm$0.081	&15.5943$\pm$0.0013	&1.1& \checkmark&\checkmark&-\\
     bd\_c7 &318.109692	&67.460453	&2221932649766967296	&309$^{+3}_{-2}$	&7.317$\pm$0.060	&-1.724$\pm$0.050	&8.1432$\pm$0.0003	&0.9&-&\checkmark&\checkmark\\
     bd\_c8 &318.504950	&68.822118	&2270438498579303808	&343$^{+5}_{-6}$	&7.950$\pm$0.078	&-1.046$\pm$0.082	&15.1636$\pm$0.0008	&1.3& \checkmark&\checkmark& \checkmark\\
     bd\_c9 &318.867019	&69.240512	&2270593529718918016	&344$^{+4}_{-4}$	&7.955$\pm$0.057	&-0.822$\pm$0.066	&15.2133$\pm$0.0041	&1.1& \checkmark&\checkmark& \checkmark\\
     bd\_c10 &318.894184	&68.757180	&2270439254493609088	&345$^{+4}_{-4}$	&7.781$\pm$0.058	&-1.172$\pm$0.060	&14.9226$\pm$0.0065	&1.1& \checkmark&\checkmark& \checkmark\\
     bd\_c11 &318.903254	&68.809839	&2270440491444181504	&348$^{+44}_{-36}$	&8.482$\pm$0.561	&-1.994$\pm$0.559	&19.5934$\pm$0.0049	&0.9&\checkmark&-&-\\            
     bd\_c12 &318.926286	&68.893498	&2270445439246710400	&326$^{+4}_{-4}$	&8.058$\pm$0.079	&-0.406$\pm$0.081	&15.0225$\pm$0.0034	&1.3& \checkmark&\checkmark& \checkmark\\
     bd\_c13 &318.926976	&68.893279	&2270445439244424576	&329$^{+11}_{-10}$	&7.966$\pm$0.213	&-1.919$\pm$0.243	&16.5337$\pm$0.0039	&-& -&\checkmark& -\\    
     bd\_c14 &318.977864	&68.681023	&2270427056786514432	&361$^{+7}_{-6}$	&7.897$\pm$0.089	&-1.130$\pm$0.087	&15.9845$\pm$0.0027	&1.1& \checkmark&\checkmark& \checkmark\\
     bd\_c15 &319.066179	&67.862292	&2222325450295259392	&340$^{+9}_{-8}$	&6.749$\pm$0.127	&-0.817$\pm$0.141	&17.2564$\pm$0.0011	&1.0&\checkmark &\checkmark&-\\
     bd\_c16 &319.251660	&69.609775	&2270618131291563392	&336$^{+11}_{-11}$	&8.150$\pm$0.179	&-0.832$\pm$0.244	&17.4985$\pm$0.0015	&1.0& \checkmark&\checkmark&-\\
     bd\_c17 &319.386343	&69.300456	&2270597481088870784	&333$^{+4}_{-4}$	&8.038$\pm$0.075	&-0.827$\pm$0.080	&15.0735$\pm$0.0017	&1.1&\checkmark&-& \checkmark\\
     bd\_c18 &319.481172	&68.990355	&2270537970021813248	&337$^{+3}_{-2}$	&7.853$\pm$0.040	&-0.439$\pm$0.032	&14.1669$\pm$0.0030	&1.1&-&\checkmark&\checkmark\\
     bd\_c19 &319.482031	&68.989987	&2270537970021813376	&349$^{+2}_{-2}$	&8.044$\pm$0.035	&-1.384$\pm$0.030	&14.0933$\pm$0.0029	&1.2&-&\checkmark&-\\
     bd\_c20  &319.525646	&68.958045	&2270525944113390848	&352$^{+6}_{-7}$	&8.417$\pm$0.087	&-1.028$\pm$0.086	&15.7204$\pm$0.0011	&1.1&\checkmark &\checkmark&\checkmark\\
     bd\_c21 &319.529989	&69.162791	&2270548758979644672	&351$^{+12}_{-11}$	&7.890$\pm$0.158	&-1.078$\pm$0.161	&16.8664$\pm$0.0014	&1.1&\checkmark &\checkmark&-\\ 
     bd\_c22 &319.548402	&68.916756	&2270524157407004032	&348$^{+6}_{-6}$	&7.999$\pm$0.087	&-1.108$\pm$0.079	&15.8677$\pm$0.0011	&1.1&\checkmark &\checkmark&-\\
     bd\_c23 &319.666806	&69.622424	&2270629500068006400	&336$^{+12}_{-11}$ 	&8.679$\pm$0.178	&-0.752$\pm$0.226	&17.5128$\pm$0.0014	&1.0& \checkmark&\checkmark&-\\
     bd\_c24 &319.772666	&68.479597	&2222380318501436800	&335$^{+21}_{-19}$	&8.712$\pm$0.323	&-1.557$\pm$0.376	&18.2628$\pm$0.0074	&1.1&\checkmark &\checkmark&\checkmark\\      
     bd\_c25 &319.778646	&69.246885	&2270550889283700224	&335$^{+4}_{-3}$	&7.972$\pm$0.062	&-0.552$\pm$0.058	&9.3328$\pm$0.0004	&1.1&-&\checkmark&\checkmark\\
     bd\_c26 &319.833473	&69.110315	&2270544837671634560	&332$^{+8}_{-8}$	&7.954$\pm$0.114	&-0.784$\pm$0.119	&16.7284$\pm$0.0013	&1.0&\checkmark &\checkmark&-\\
     bd\_c27 &319.890995	&69.078856	&2270533022219508352	&340$^{+5}_{-4}$	&8.000$\pm$0.061	&-0.817$\pm$0.061	&15.6177$\pm$0.0019	&1.0&\checkmark &\checkmark&\checkmark\\
     bd\_c28 &319.955372	&68.675289	&2222478797807817600	&339$^{+4}_{-3}$	&8.150$\pm$0.050	&-1.304$\pm$0.051	&15.0690$\pm$0.0008	&1.0&\checkmark &\checkmark&\checkmark\\
     bd\_c29 &320.131710	&68.861316	&2222488212376118912	&330$^{+12}_{-11}$	&8.020$\pm$0.179	&-1.905$\pm$0.212	&17.2170$\pm$0.0011	&1.1&\checkmark &\checkmark&-\\      
     bd\_c30 &320.251722	&69.046986	&2270530200423580416	&310$^{+22}_{-19}$	&7.637$\pm$0.294	&-1.360$\pm$0.350	&18.3472$\pm$0.0020	&1.1&\checkmark &\checkmark&-\\
     bd\_c31 &320.344413	&68.078359	&2222308339145993728	&333$^{+38}_{-31}$	&9.425$\pm$0.494	&-2.403$\pm$0.528	&19.0752$\pm$0.0031	&1.0&\checkmark &-&-\\
     bd\_c32 &320.684040	&69.461889	&2270579614025029376	&360$^{+19}_{-17}$	&8.633$\pm$0.245	&-0.981$\pm$0.269	&17.7769$\pm$0.0038	&1.1&-&\checkmark&\checkmark\\
     bd\_c33 &320.982123	&69.848891	&2272089896325294720	&357$^{+2}_{-2}$    &6.139$\pm$0.030	&-0.584$\pm$0.032	&14.5512$\pm$0.0006	&1.1&\checkmark &\checkmark&-\\      
     bd\_c34 &321.319431	&69.870931	&2272090892757724416	&336$^{+7}_{-7}$	&8.424$\pm$0.114	&-0.801$\pm$0.123	&16.1774$\pm$0.0029	&1.2&\checkmark&\checkmark&-\\
     bd\_c35 &321.726628	&68.853701	&2222452856205683968	&362$^{+10}_{-10}$	&8.404$\pm$0.153	&-1.560$\pm$0.140	&16.9416$\pm$0.0020	&1.0&\checkmark &\checkmark&-\\
     bd\_c36 &321.853223	&69.272097	&2224021893658285952	&364$^{+14}_{-14}$	&8.627$\pm$0.177	&-0.859$\pm$0.205	&17.3336$\pm$0.0013	&1.1&\checkmark &\checkmark&-\\
     bd\_c37 &322.245905	&69.633510	&2224035946791401216	&340$^{+5}_{-4}$	&8.498$\pm$0.076	&-1.076$\pm$0.076	&15.4698$\pm$0.0022	&1.1&\checkmark &\checkmark&\checkmark\\
     bd\_c38 &323.560526	&69.634333	&2224048178858580736	&333$^{+6}_{-5}$	&8.709$\pm$0.087	&-0.699$\pm$0.089	&14.7469$\pm$0.0027	&1.3&\checkmark &\checkmark&\checkmark\\
     bd\_c39 &323.869465	&69.568596	&2223997326445821696	&347$^{+8}_{-7}$	&8.730$\pm$0.095	&-1.067$\pm$0.087	&16.2078$\pm$0.0013	&1.1&\checkmark &\checkmark&-\\
      \hline
    \end{tabular}\\
\end{table*}    
\begin{table*}
    \begin{center}
    \caption*{Table \ref{tab:YSO_new} continued.}
    \renewcommand{\arraystretch}{1.3}
    \tiny
    \begin{tabular}{p{0.7cm}p{1.2cm}p{1.1cm}p{2.6cm}p{0.8cm}p{1.3cm}p{1.6cm}p{1.8cm}p{0.8cm}p{0.8cm}p{0.8cm}p{0.8cm}}
    \hline
    $\#$& RA(2015.5) & Dec(2015.5) & Source & Distance & $\mu_{\alpha\star}$ ($\Delta\mu_{\alpha\star}$) & $\mu_{\delta}$ ($\Delta\mu_{\delta}$) & G (eG) & RUWE &Pan- &2MASS & $WISE$\\
     & ($^{\circ}$) & ($^{\circ}$) & ID & (pc) & (mas yr$^{-1}$) & (mas yr$^{-1}$) & (mag) & & STARRS & & \\
     (1)&(2)&(3)&(4)&(5)&(6)&(7)&(8)&(9)&(10)&(11)&(12)\\
      \hline
      \multicolumn{12}{c}{sources not associated with any cloud} \\
      o\_c1&307.672240	&68.438749	&2247137609506697728	&367$^{+35}_{-30}$	&6.234$\pm$0.475	&-1.960$\pm$0.632	&19.1158$\pm$0.0036	&0.9&\checkmark &-&-\\
      o\_c2&307.929469	&66.343862	&2246345346956211200	&311$^{+14}_{-12}$	&7.575$\pm$0.280	&-1.698$\pm$0.331	&18.3046$\pm$0.0025	&1.0&\checkmark &-&-\\
      o\_c3&309.820816	&66.616681	&2245991235492832512	  &341$^{+12}_{-10}$	&9.387$\pm$0.181	&-2.371$\pm$0.178	&17.5811$\pm$0.0016	&0.9& \checkmark &\checkmark&-\\
      o\_c4&310.902769	&70.134329	&2271441184463842560	&343$^{+2}_{-2}$	&7.855$\pm$0.035	&-2.885$\pm$0.035	&10.1049$\pm$0.0002 &1.0&-&\checkmark&\checkmark\\
      o\_c5&311.509466	&71.085859	&2274882930735828736	&340$^{+9}_{-8}$	&6.938$\pm$0.115	&-0.747$\pm$0.106	&16.8358$\pm$0.0010	&1.0&\checkmark &\checkmark&\checkmark\\
      o\_c6&311.520948	&66.124230	&2245924440161432064	&365$^{+2}_{-3}$	&7.450$\pm$0.034	&-0.779$\pm$0.031	&13.8946$\pm$0.0020	&1.1&\checkmark &\checkmark&-\\
      o\_c7&311.733175	&66.018168	&2245735113706518528	&370$^{+15}_{-14}$	&7.248$\pm$0.197	&-0.956$\pm$0.190	&17.5133$\pm$0.0016	&1.1&\checkmark &\checkmark&-\\
      o\_c8&312.080716 & 67.554984 &2246149767027696896   &339$^{+15}_{-14}$ & 7.872$\pm$0.247 & -1.072$\pm$0.220 & 17.9527$\pm$0.0017 & 1.0& \checkmark &\checkmark&-\\
      o\_c9&312.957679	&68.947681	& 2271099648664907392	&341$^{+29}_{-24}$	&9.503$\pm$0.418	&-1.957$\pm$0.493	&19.0595$\pm$0.0030	&0.9&\checkmark &-&-\\     
      o\_c10&313.117760	&69.422860	&2271163523418749568	&321$^{+3}_{-2}$	&8.575$\pm$0.045	&-1.681$\pm$0.051	&11.3718$\pm$0.0003	&0.9&\checkmark &\checkmark&\checkmark\\      
      o\_c11&314.025942	&66.036337	&2245760269331745664	&373$^{+3}_{-3}$	&8.967$\pm$0.041	&-1.109$\pm$0.038	&14.5641$\pm$0.0004	&0.8&\checkmark &\checkmark&-\\
      o\_c12&314.679887	&70.192278	&2271586045120993792	&314$^{+2}_{-2}$	&8.082$\pm$0.034	&-1.173$\pm$0.032	&14.3251$\pm$0.0004	&0.9&\checkmark &\checkmark&\checkmark\\      
      o\_c13&314.700118	&66.040266	&2197740915758258816	&332$^{+22}_{-20}$	&8.761$\pm$0.343	&-1.131$\pm$0.342	&18.5524$\pm$0.0019	&1.1&\checkmark &-&-\\
      o\_c14&314.848513	&65.534143	&2197648354919811456	&358$^{+30}_{-25}$	&7.482$\pm$0.389	&-1.710$\pm$0.398	&18.8231$\pm$0.0028	&1.1&\checkmark &-&-\\
      o\_c15&315.812443	&66.704797	&2221785933684340736	&329$^{+8}_{-8}$	&7.431$\pm$0.155	&-2.773$\pm$0.147	&16.7894$\pm$0.0015	&1.0&\checkmark &\checkmark&\checkmark\\      
      o\_c16&316.373414	&64.796710	&2197367528481936128	&333$^{+25}_{-22}$	&7.795$\pm$0.413	&-2.264$\pm$0.344	&18.8689$\pm$0.0028	&1.0 &\checkmark &-&-\\    
      o\_c17&321.967321	&70.492754	&2272185072800774784	&360$^{+8}_{-8}$	&8.798$\pm$0.098	&-0.391$\pm$0.105	&16.2997$\pm$0.0027	&1.1&\checkmark &\checkmark&\checkmark\\
      o\_c18&321.977437	&67.034826	&2221257304812229376	&351$^{+17}_{-15}$	&8.938$\pm$0.228	&-1.404$\pm$0.281	&18.1570$\pm$0.0016	&1.0&\checkmark &\checkmark&-\\
    \hline
    \end{tabular}\\
    \end{center}
    \tiny
    \renewcommand{\arraystretch}{1.0}
\end{table*}  
\end{appendix}

\end{document}